\documentclass[aps,prd,10pt,onecolumn,nofootinbib,showpacs,showkeys,notitlepage]{revtex4-1}
\usepackage{graphicx,bm}
\usepackage{hyperref}
\usepackage{amssymb}
\usepackage{caption}
\usepackage{subcaption}
\usepackage{amsmath}
\usepackage{color}

\newcommand{\cD}{{\cal D}}

\newcommand{\nn}{\nonumber\\}

\newcommand{\rh}{\varrho}

\newcommand{\exv}[1]{\left\langle{#1}\right\rangle}

\newcommand{\ep}{\varepsilon}

\newcommand{\p}{{\bf p}}

\newcommand{\sgn}{\mathop{\textrm{sgn}}}
\newcommand{\Disc}{\mathop{\textrm{Disc}}}

\newcommand{\pint}[2]{{\int\!\frac{d^{#1}#2}{(2\pi)^#1}\,}}

\newcommand{\G}{{\cal G}}

\begin{document}

\title{Validating the 2PI resummation: the Bloch-Nordsieck example}

\author{A. Jakov\'ac}
\email{jakovac@caesar.elte.hu}
\affiliation{Institute of Physics, E\"otv\"os University,
H-1117 
Budapest, Hungary}
\author{P. Mati}
\email{mati@phy.bme.hu, matipeti@gmail.com}
\affiliation{Institute of Physics, Budapest University of Technology
  and Economics, H-1111 Budapest, Hungary}
\affiliation{Institute of Physics, E\"otv\"os University,
H-1117 
Budapest, Hungary}
\affiliation{ MTA-DE Particle Physics Research Group,
  University of Debrecen, H-4010 Debrecen P.O.Box 105, Hungary}

\begin{abstract}
  In this work we provide a numerical method to obtain the
  Bloch-Nordsieck spectral function at finite temperature in the
  framework of the 2PI approximation. We find that the 2PI results
  nicely agree with the exact one, provided we perform a coupling
  constant matching. In the paper we present the resulting finite
  temperature running of the 2PI coupling constant. This result may
  apply for the finite temperature behavior of the coupling constant
  in QED, too.
\end{abstract}

\maketitle

\section{Introduction}

The infrared limit of the QED was modeled by Bloch and Nordsieck in
1937, and their treatment of the IR singularities has become a
textbook material since. In the framework of the Bloch-Nordsieck (B-N)
model one is able to resum all of the radiative contributions to the
fermionic Green's function generated by ultra-soft photons. A detailed
discussion of this calculation can be found in the original paper of
Bloch and Nordsieck \cite{BN}, in textbooks
\cite{BogoljubovShirkov},\cite{Fried} and in \cite{Jakovac:2011aa}.

Besides an exact solution one can also give solutions in different
approximations. In particular the 2-particle-irreducible (2PI)
approximation \cite{2PIhist} is a well known tool to study the quasiparticle
properties of the system. One of the biggest challenge in front of the
2PI techniques is the representations of symmetries, in particular
gauge symmetries \cite{RS1}. The study of B-N model provides an
excellent tool to test the reliability of fixed gauge calculations.

In our paper \cite{Jakovac:2011aa} we used the 2PI functional method
to study the B-N model at zero temperature. The spectrum could be
obtained by applying numerical calculations. We found on the one hand
the disappointing fact that the fit comparison to the exact propagator
was not very promising (see \cite{Jakovac:2011aa}), on the other hand,
unlike in the old-fashioned perturbation theory, the spectrum remained
regular even in the highly infrared regime (no infrared singularity
observed at the mass shell).

At finite temperature there are several studies in the literature
\cite{IancuBlaizot1, IancuBlaizot2,Fried:2008tb} to derive the
behavior of the fermion propagator. In our paper \cite{Jakovac:2011aa}
we developed a method to reproduce the exact result using the
Ward-Takahashi identities at zero temperature (cf. also
\cite{Alekseev:1982dk}).
This could have been generalized to finite temperature in
\cite{Jakovac:2013aa}. With the help of this method we managed to
obtain a fully analytical form of the excitation spectrum. Having
these analytic results gives us a perfect opportunity to investigate
the validity of the 2PI quasiparticle description of an interacting
quantum field theory at finite temperature.

The purpose of the present paper is to show how the 2PI works at
finite temperature. We will derive the spectral function numerically
and compare it to the exactly calculated case. The upshot is: there
exists a mapping between the coupling constants of the 2PI and the
exact results in such a way that the two spectral function overlap
almost entirely. This is a highly nontrivial result, since the exact
spectral function is an asymmetric function of the frequency, rather
different from a simple Lorentzian. The most important message to the
2PI community is that our result validates the 2PI approximation
method at non-zero temperature, and only a finite reparametrization of
the theory is needed.

From the perturbative point of view the 2PI technique resums the two
particle irreducible diagrams, but the coupling constant and also the
higher point functions are remained unchanged. So for a certain 2PI
diagram there exists another infinite set of diagrams providing
coupling constant modification. In the sense of the renormalization
group we may try to take into account the sum of these diagrams
effectively as a temperature dependent (running) coupling
constant. Since we now know the value of an observable exactly (the
electron spectral function for any frequency and temperature in a
given gauge), the best method to extract the temperature dependent
coupling is to compare the 2PI and the exact results. This is done in
the present paper.

The structure of the paper is as follows. First we give an
introduction to the Bloch-Nordsieck model itself and to the
conventions of the finite temperature real time formalism. In Section
III we recap the zero temperature results: the one-loop correction
obtained from perturbation theory (PT) and the implementation of the
2PI numerics. In Section IV we derive the one-loop self energy at
$T\neq 0$ and show its consistency with the zero temperature result by
taking the $T\rightarrow 0$ limit. Then we calculate the expression
for the discontinuity of the self energy for the 2PI procedure. The
numerical implementation of the calculation happens in a similar
fashion as the zero temperature one. In Section V we present our
results obtained from the numerics and the comparison to the exact
result \cite{Jakovac:2013aa}. We found a non-trivial mapping of the
coupling between the two calculation methods from which we conclude:
the 2PI, although it is an approximation, at finite temperature it
gives a perfect qualitative description of the collective excitation
of the system.

\section{The properties of the Bloch-Nordsieck model}

The Bloch-Nordsieck model was designed to describe accurately the low
energy regime of quantum electrodynamics. Considering the
contributions to the fermion self energy only from the deep infrared
photons, reducing the Dirac spinor to a one component fermion is well
justified. Indeed, at this energy scale photons do not have enough
energy even to flip the spin of the fermion, not to mention the pair
creation \cite{BN}. In this respect, one can substitute a four vector
$u_\mu$ in the place of the $\gamma_\mu$ matrices which is considered
as the four velocity of the fermion.

The Lagrangian then reads:
\begin{eqnarray}
\mathcal{L}=-\frac{1}{4}F_{\mu\nu}F^{\mu\nu}+\Psi^{\dagger}(iu_{\mu}D^{\mu}-m)\Psi. 
\end{eqnarray} 
Where the usual notations for the field-strength tensor and for the covariant derivative are used: $F_{\mu\nu}=\partial_{\mu}A_\nu-\partial_{\nu}A_\mu$ and $D_{\mu}=i\partial_{\mu}-eA_{\mu}$, respectively. 
Here, as it was mentioned above, $u_{\mu}$ is a four velocity, but we can also choose the form $u=(1,\bf{v})$, with ${\bf{v}}={\bf{u}}/u_0$ by rescaling the fermionic field as $\Psi\rightarrow\Psi/\sqrt{u_0}$ and the fermion mass by $m\rightarrow m u_0$.
\par It is possible to obtain exactly the full fermion propagator associated with this theory both for zero and finite temperatures as it is presented in \cite{Jakovac:2011aa} and \cite{Jakovac:2013aa}, respectively. Now we are going to discuss the notations and conventions that we are using in this paper.\\
For the calculations we used the real time formalism (details in
\cite{Jakovac:2013aa},\cite{LeBellac}). The propagators are matrices
in this convention: 
\begin{equation}
  i{\cal G}_{ab}(x)=\exv{T_C \Psi_a(x) \Psi_b^\dagger (0)}\qquad
  \mathrm{and}\qquad iG_{\mu\nu,ab}(x)=\exv{T_C A_{\mu a}(x) A_{\nu b} (0)},
\end{equation}
where $T_C$ denotes ordering with respect to the contour variable
(contour time ordering). At finite temperature with help of the KMS relation we can determine $G_{12}$ and $G_{21}$
\begin{equation}
  \label{eq:id1}
  iG_{12}(k) = \pm n_\pm(k_0) \rh(k),\qquad iG_{21}(k) = (1\pm
  n_\pm)(k_0) \rh(k),
\end{equation}
where 
\begin{equation}
  n_\pm(k_0) = \frac1{e^{\beta k_0}\mp 1}\quad \mathrm{and}\quad \rh(k)
  = iG_{21}(k)-iG_{12}(k)
\end{equation}
are the distribution functions (Bose-Einstein (+) and Fermi-Dirac (-)
statistics), and the spectral function, respectively. We will also use
R/A formalism \cite{Jakovac:2013aa}, where
\begin{equation}
  \label{eq:id2}
  G_{rr}=\frac{G_{21}+G_{12}}2,\quad G_{ra} = G_{11} - G_{12},\quad
  \rh=i G_{ra}-iG_{ar}.
\end{equation}
The $G_{ra}$ propagator is the retarded, the $G_{ar}$ is the advanced propagator, $G_{rr}$ is usually called the Keldysh propagator.

At zero temperature the (free) fermionic Feynman-propagator reads:
\begin{equation}\label{ferm}
  {\cal G}_0(p) = \frac1{u_\mu p^\mu -m +i\ep}.
\end{equation}
It has a single pole which means that there are no antiparticles in
the model. Consequently, closed fermion loops are excluded, thus there
is no self energy correction to the photon propagator at zero
temperature. Physically this means that the energy is not sufficient
to excite the antiparticles. We interpret $u^\mu$ as the fermionic
four-velocity, and since it is fixed, the soft photons cannot change
it (no fermion recoil). In fact this means that the fermion is a hard
probe of the system, hence not part of the thermal medium
\cite{IancuBlaizot1, IancuBlaizot2}. This is also supported by
the spin-statistics theorem \cite{PeshkinSchroeder} which would forbid
a one-component fermion field. Consequently we will neglect the '12'
fermion propagator, too: ${\cal G}_{12}=0$.

The exact photon propagator reads in Feynman gauge
\begin{equation}\label{phot}
  G_{ab,\mu\nu}(k) = -g_{\mu\nu} G_{ab}(k),\qquad G_{ra} =
  \frac1{k^2}\biggr|_{k_0\to k_0+i\ep},\quad \rh(k)= 2\pi\sgn(k_0) \delta(k^2),
\end{equation}
all other propagators can be expressed using identities \eqref{eq:id1}
and \eqref{eq:id2}. 

\section{Recap of T=0 2PI calculations}

The main idea is to use the exact propagators in the perturbation
theory as building blocks of a loop integral, where the exact
propagator is determined self-consistently using skeleton diagrams as
resummation patterns. The one-loop 2PI fermion self energy diagram in the
case of the Bloch-Nordsieck model generates the resummation of all the
``rainbow'' diagrams. One needs to take care of the UV divergences,
too, on which we perform a renormalization with the same form of
divergent parts of the counterterms as in the 1-loop case.

At zero temperature we have the following self-consistent system of
equations in the 2PI approximation
\begin{eqnarray}\label{self}
\G(p)&=&\frac{1}{\G_{0}^{-1}(p)-\Sigma(p)}, \\
\label{selfloop}
-i\Sigma(p)&=&(-ie)^2\pint{4}{k}i G_{\mu\nu}(k)i\G(p-k). 
\end{eqnarray}
Here $\G_0$ and $\G$ stand for the free and the dressed fermion propagator. $G_{\mu\nu}$ is the photon propagator.

\subsection{The one-loop correction}
In strict perturbation theory (PT) we use the propagators from Eq.(\ref{ferm}) and (\ref{phot}) to compute the self energy. We choose a reference frame which in $u^{\mu}=(1,0,0,0)$, and we find using dimensional regularization
\begin{equation}
\Sigma_{1loop}(p_0)=\frac{\alpha}{\pi}(p_0-m)\left(-\ln{\frac{m-p_0}{\mu}} + \cD_\epsilon \right).  
\end{equation} 
Here $\mu$ is the renormalization scale. The divergent part $\cD_\epsilon$ has the following expression:
\begin{equation}
\cD_\epsilon=\frac{1}{2\epsilon}+\frac{1}{2}\left(\ln{4\pi} -\gamma_E \right).
\end{equation}
We renormalized the self energy using the $\overline{\text{MS}}$ scheme, by which the counterterms read as

\begin{equation}\delta Z_{1,\overline{\text{MS}}}=\delta_{m,\overline{\text{MS}}}= \frac{\alpha}{\pi}D_\epsilon,
\end{equation}

where $\delta Z_{1,\overline{\text{MS}}}$, $\delta_{m,\overline{\text{MS}}}$ are the wave function renormalization and the multiplicative mass renormalization, respectively.
Hence, the renormalized self energy is 
\begin{equation}\label{SE_1LO}
\Sigma_{1loop}^{ren}=-\frac{\alpha}{\pi}(p_0-m)\ln{\frac{m-p_0}{\mu}}.
\end{equation}
For the details see \cite{Jakovac:2011aa}.

\subsection{The 2PI procedure at $T=0$}
In the 2PI approach we treat Eq.(\ref{self}) and Eq.(\ref{selfloop}) self-consistently. Then we implement the following steps numerically \cite{Jakovac:2011aa}, which will be applied at finite T, too:\newline
\vskip 5pt \textit{step 1:} We calculate the discontinuity of the self-energy in order to use it the spectral representation of the retarded Green's function.
\begin{eqnarray}
\Sigma(p)=i e^2\pint{4}{k}G_{\mu\nu}\G(p-k)&=&ie^2\int\limits_{0}^{\infty}\frac{d\omega}{2\pi}\pint{4}{k}\frac{1}{k^2+i\epsilon}\frac{\rho(\omega)}{p_0-k_0-\omega+i\epsilon}\\
\Sigma(p)&=&\int\limits_{0}^{\infty}\frac{d\omega}{2\pi}\rho(\omega)\Sigma_{1-loop}(p,\omega)
\end{eqnarray}
Now we can take the discontinuity
\begin{eqnarray}
\Disc\limits_{p_0}\Sigma(p)=\int\limits_{0}^{\infty}\frac{d\omega}{2\pi}\rho(\omega)\Disc\limits_{p_0}\Sigma_{1-loop}(p,\omega)=
\frac{\alpha}{\pi}\int\limits_{0}^{\infty}d\omega(p_0-\omega)\rho(\omega).
\end{eqnarray}
In both equations we introduced the fermion spectral function $\rho(p)$. In our algorithm it serves as an input, which is usually the free fermion spectral function $\rho(p)=2\pi\delta(p-m)$.\newline
\vskip 5pt \textit{step 2:} Here we calculate the real part of self energy from its discontinuity. For this purpose we use the Kramers-Kronig relation:
\begin{eqnarray}
\text{Re}\Sigma(p_0,{\bf{\p}})=\int\limits_{-\infty}^{\infty}\frac{d \omega}{2 \pi} \frac{\Disc_{\omega}i\Sigma(\omega,{\bf p })}{p_0-\omega+i\epsilon}.
\end{eqnarray}
\textit{step 3:} We renormalize the real part of the self energy using the "on-mass-shell" (OMS) renormalization scheme:
\begin{eqnarray}
&\text{Re}\Sigma(p_0=m)&=0,\\
&\displaystyle\left.\frac{d \text{Re}\Sigma(p_0)}{d p_0}\right|_{p_0=m}&=0.\\
\end{eqnarray} 
\vskip 5pt \textit{step 4:} From all of this information we construct the new spectral function, which reads as
\begin{eqnarray}
\rho(p)=\frac{2\text{Im}\Sigma(p)}{(p_0-m-\text{Re}\Sigma(p))^2+(\text{Im}\Sigma(p))^2}.
\end{eqnarray}
\vskip 5pt \textit{step 5:} We set the new spectral function to be our new input, and iterate this procedure till it converges.\newline
\vskip 5pt \textit{step 4+:} As an additional step we had to include a rescaling of the spectral function which was necessary to stabilize the convergence. This step is not needed at non-zero temperature.\newline
For the zero temperature case we obtained the dressed propagator for the fermion. From the analysis it turned out that this result, being an approximation, is far from the exact solution although it is infrared finite, which cannot be claimed about the PT calculation (see \cite{Jakovac:2011aa}).

\section{Non-zero temperature}

\begin{figure}[h!]
\centering
 \includegraphics[scale=0.18]{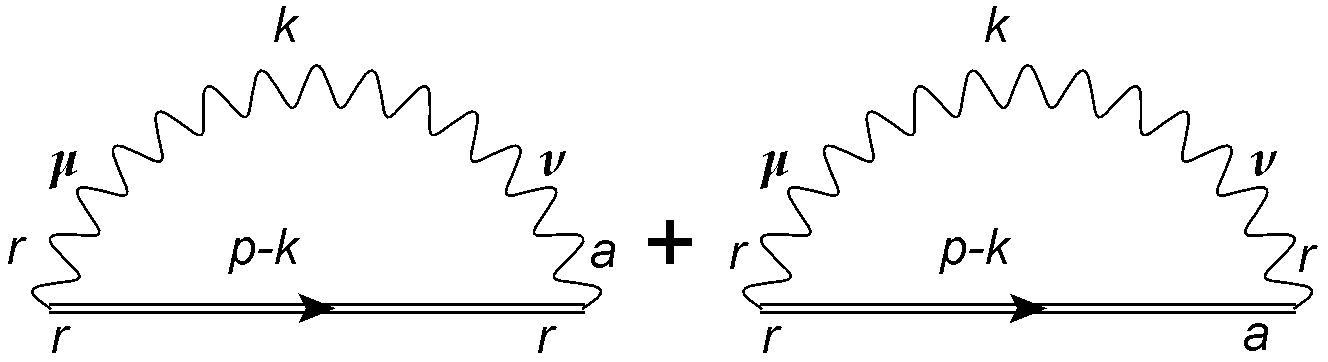}
 \caption{The diagrammatic representation of the self energy. The wavy line corresponds to the free (here also the exact) photon propagator with a loop momentum $k$ and the double solid line is for the exact fermion propagator with momentum $p-k$. Both the polarization and the Keldysh indices are shown.}
 \label{diag}
\end{figure}

We are working in the real time formalism, hence the Green's functions
in this picture are going to have a matrix structure. We choose the
$R/A$ basis for the matrix representation to calculate the retarded
self energy. First we are going to consider the one-loop correction
then we present a derivation of the 2PI resummed spectral function at
finite temperature. To evaluate its self-consistent equations we will
use a numerical approach which is similar to what we discussed above
for the $T=0$ case.  The integral equation for the retarded self
energy at non-zero temperature in Feynman gauge reads as:

\begin{eqnarray}\label{selfT}
\Sigma_{ar}(p)=ie^2\pint{4}{k} \left[ G_{rr}(k)\G_{ra}(p-k)+G_{ra}(k)\G_{rr}(p-k) \right].\end{eqnarray}
Where $G$ and $\G$ stands for the propagator of the photon and the fermion, respectively. On Fig.(\ref{diag}) we can see the pictorial representation of the fermion self energy using Feynman diagrams.

Now, if we take the discontinuity we will have:
\begin{eqnarray}\label{sigmad}
\Disc\limits_{p_0}\Sigma_{ar}(p)=e^2\pint{4}{k}  \left[ G_{rr}(k)\rho_{f}(p-k)+\rho_{\gamma}(k)\G_{rr}(p-k)\right].
\end{eqnarray}
Here $\rho_f$ and $\rho_\gamma$ are the spectral functions to the fermion and the photon, respectively.
In general we can express the $rr$ propagators by the spectral function combining with the distribution function of the corresponding spin statistics:
\begin{eqnarray}
\G_{rr}(p)&=&\left(\frac{1}{2}-n_{f}(p_0)\right)\rho_{f}(p),\\
G_{rr}(p)&=&\left(\frac{1}{2}+n_{b}(p_0)\right)\rho_{\gamma}(p).
\end{eqnarray}
When inserting these expressions into Eq.(\ref{sigmad}), we get
\begin{eqnarray}\label{general}
\Disc\limits_{p_0}\Sigma_{ar}(p)&=&e^2\pint{4}{k}  \left[ \left(\frac{1}{2}+n_{b}(k_0)\right)\rho_{\gamma}(k)\rho_{f}(p-k)+\rho_{\gamma}(k)\left(\frac{1}{2}-n_{f}(p_0-k_0)\right)\rho_{f}(p-k)\right]=\nn
&=& e^2\pint{4}{k} \left(1+n_{b}(k_0)-n_{f}(p_0-k_0)\right)\rho_{\gamma}(k)\rho_{f}(p-k). 
\end{eqnarray}
In the last step of Eq.(\ref{general}) we get the most general form of the equation, as long as we do not specify the corresponding spectral functions.

\subsection{One-loop correction at $T\neq 0$}

For the one-loop case we have to plug in the spectral function of the free theory both for the fermion and gauge fields.
By performing this substitution our equation reads as
\begin{eqnarray}\label{long}
\Disc\limits_{p_0}\Sigma_{ar}(p)&=& e^2\pint{4}{k} \left(1+n_{b}(k_0)-n_{f}(p_0-k_0)\right)2\pi\sgn {k_0}\delta(k_{0}^2-{\bf k}^2)2\pi\delta\left(u_0(p_0-k_0)-{\bf u}({\bf p}-{\bf k})-m\right)=\nn
&=&\frac{e^2}{8\pi^3}\int\limits_{0}^{\infty} \! d k {\bf k}^2 \int\limits_{-1}^{1} \! dx \frac{(2\pi)^2}{2|{\bf k}|} \left[\left(1+n_{b}(|{\bf k}|)-n_{f}(p_0-|{\bf k}|)\right)
\delta\left(u_0p_0-{\bf u}{\bf p} -u_0|{\bf k}| -|{\bf u}||{\bf k}|x-m\right)+\right.\nn
&+&\left. \left(n_{b}(|{\bf k}|)+n_{f}(p_0+|{\bf k}|)\right)\delta\left(u_0p_0-{\bf u}{\bf p} +u_0|{\bf k}| -|{\bf u}||{\bf k}|x-m\right)\right].
\end{eqnarray}
Here we introduced the variable $x$ which stands for the cosine of the angle between the two spatial three-vectors ${\bf u}$ and ${\bf k}$. For the sake of simplicity in the following we are going to use the notations $pu\equiv p_0u_0-{\bf{pu}}$ for the Minkowski product, and $k\equiv |{\bf k}|,u\equiv|{\bf u}|$ for the absolute values of the three vectors ${\bf k}$ and ${\bf u}$, respectively.\\
First we perform the angular integration for $x$: 

\begin{eqnarray}\label{asd}
\Disc\limits_{p_0}\Sigma_{ar}(p)=\frac{e^2}{4\pi u} \left(\Theta(pu-m)\int\limits_{\frac{pu-m}{u+u_0}}^{\frac{pu-m}{u-u_0}}  \! dk \left(1+n_{b}(k)-n_{f}(p_0-k)\right)+\Theta(m-pu)\int\limits_{\frac{m-pu}{u+u_0}}^{\frac{m-pu}{u-u_0}}  \! dk \left(n_{b}(k)+n_{f}(p_0+k)\right)\right)
\end{eqnarray}

Now we are going to use the fact that the fermion in this system is a
hard probe, thus it is not part of the heat-bath \cite{IancuBlaizot1,
  IancuBlaizot2}. This manifests already in Eq.(\ref{general}) in a
way that we need to set the Fermi-Dirac distribution to zero,
otherwise we would face with inconsistencies when one would try to
take the $T\rightarrow 0$ limit:
\begin{equation}\label{nofer} n_f\equiv 0 \text{    (in the B-N framework).}\end{equation}
In that case Eq.(\ref{asd}) simplifies in the following way:
\begin{eqnarray}\label{asd2}
\Disc\limits_{p_0}\Sigma_{ar}(p)=\frac{e^2}{4\pi u} \int\limits_{\frac{pu-m}{u+u_0}}^{\frac{pu-m}{u-u_0}}  \! dk \left(1+n_{b}(k)\right).
\end{eqnarray}
Evaluating the integral one gets a result consistent with the $T=0$ case:
\begin{equation}
\Disc\limits_{p_0}\Sigma_{ar}(p)=\frac{e^2}{2\pi}\Theta(pu-m)(pu-m)+\frac{Te^2}{4\pi u}\ln\left(\frac{1-e^{-\beta \frac{pu-m}{u-u_0}}}{1-e^{-\beta \frac{pu-m}{u+u_0}}}\right).
\end{equation}
This gives us the desired result for the $T\rightarrow 0$ limit, namely $\Disc\limits_{p_0}\Sigma_{ar}(p)=\frac{e^2}{2\pi}\Theta(pu-m)(pu-m)$.

\subsection{Non-zero temperature calculations for the 2PI scheme}
Now we are going to derive the 2PI resummed result for the finite temperature theory. Let us consider Eq.(\ref{self}) and Eq.(\ref{selfT}). Instead of calculating the one-loop correction by inserting free propagators, we are going to use the self-consistent fermion propagator so defining a self-consistent system of integral equations. We stick to the physical picture that the fermion is not part of the thermal medium, Eq.(\ref{nofer}) . Using the calculation in Eq.(\ref{long}) we arrive to an expression for the discontinuity of the self energy for general fermion propagator:
\begin{eqnarray}
\Disc\limits_{p_0}\Sigma_{ar}(p)&=&e^2\pint{4}{k}\left(1+n_b(k_0)\right)\rho_\gamma(k)\rho_f(p-k)\nn
&=&e^2\pint{4}{k}\left(1+n_b(k_0)\right)\frac{2\pi}{2k}(\delta(k_0-k)-\delta(k_0+k))\bar\rho_f(up-uk-m). 
\end{eqnarray}
Here we used the free photon propagator as above and for the general spectral function of the fermion we introduced the notation $\rho_f(p)=\bar\rho_f(up-m)$. After some algebra we find 
\begin{eqnarray}
\Disc\limits_{p_0}\Sigma_{ar}(p)=\frac{e^2}{8\pi^2}\int\limits_{-\infty}^{\infty}\! dk \int\limits_{-1}^{1} \! dx k n_b(k)\bar\rho_f(w+(u_0+ux)k). 
\end{eqnarray}
Here we defined $w:=up-m$, and $x$ represents the angle between the spatial parts of $k^\mu$ and $u^\mu$, so $xku$ is the scalar product of two three-dimensional vectors like in the one-loop calculation. Actually, this can be written in a more elegant, and for the numerical implementation, a more useful way. We introduce the variable $z$ as the argument of the function $\bar\rho_f$:
\begin{eqnarray}
\Disc\limits_{p_0}\Sigma_{ar}(p)&=&\frac{e^2}{8\pi^2}\frac{1}{u}\int\limits_{-\infty}^{\infty}\! dk \int\limits_{w+(u_0-u)k}^{w+(u_0+u)k}\! dz \bar\rho_f(z)  n_b(k)=\frac{e^2}{8\pi^2}\frac{1}{u}\int\limits_{-\infty}^{\infty}\! dz \bar\rho_f(z) \int\limits_{\frac{z-w}{u_0-u}}^{\frac{z-w}{u_0+u}} \! dk   n_b(k).
\end{eqnarray}
In the case when the length of the three-velocity tends to zero,$u\rightarrow 0$ , we have
\begin{eqnarray}\label{u0}
\Disc\limits_{p_0}\Sigma_{ar}(p_0)=\frac{\alpha}{\pi}\int\limits_{-\infty}^{\infty} dz \bar\rho_{f}(z)(p_0-m-z)(1+n_b (p_0-m-z)).
\end{eqnarray}   
For $u\neq 0$ 
\begin{eqnarray}\label{ufin}
\Disc\limits_{p0}\Sigma_{ar}(w)=\frac{\alpha}{2\pi}\int\limits_{-\infty}^{\infty} dz \bar\rho_f(z)\frac{T}{u}
\ln\frac{1-e^{-\beta\frac{z-w}{u_0-u}}}{1-e^{-\beta\frac{z-w}{u_0+u}}}.
\end{eqnarray}
We set $m=0$, this can be done without the loss of generality since
the two expressions in Eq.($\ref{u0}$) and Eq.($\ref{ufin}$) depends
on the variable $w=up-m$ only. That means the theory is not sensitive
where the mass-shell is placed, it can be anywhere on the real line.
With these formulae we can find the solution of the self-consistent
system of equations numerically, in a similar way we did it in the
case of zero temperature ({\it{step 1 - step 5}}), but this time we
insert the finite temperature ingredients (Eqs.~\ref{u0} and \ref{ufin}).

\section{2PI results}

We are implementing the same numerical method that we used for the
zero temperature case (Section III/B); the renormalization procedure
goes in the same way. First we observe that a small thermal mass is
generated. Interestingly this thermal mass is negative, it shifts the
spectral function to the left. In the exact solution in
\cite{Jakovac:2013aa} we found a zero thermal mass, thus we can
consider it as an artifact of the 2PI approximation, which can be
incorporated into the mass, or into the notation $w=up-m$. We should
also remark that the B-N theory describes only the softest photons
while in a realistic theory the mass modification dominantly is due to
the modes with higher momenta.

\subsection{The zero velocity case} 

\begin{figure}[htbp]
  \centering
  \includegraphics[height=6cm]{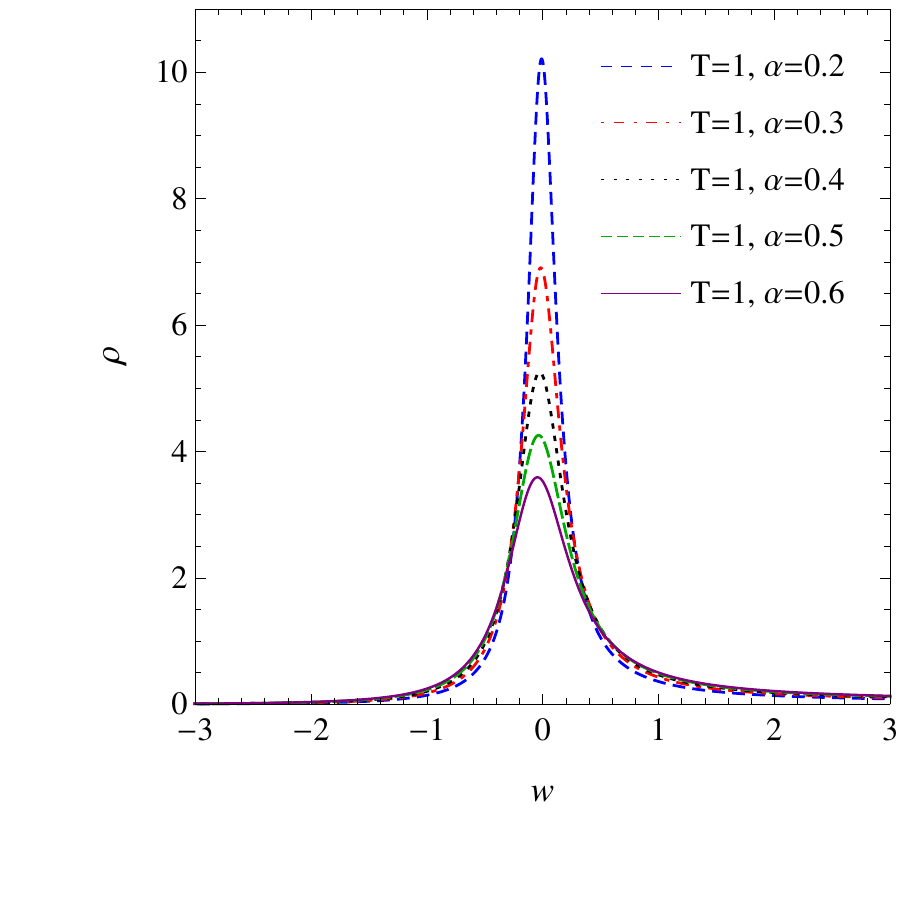}
  \hspace*{1cm}
  \includegraphics[height=6cm]{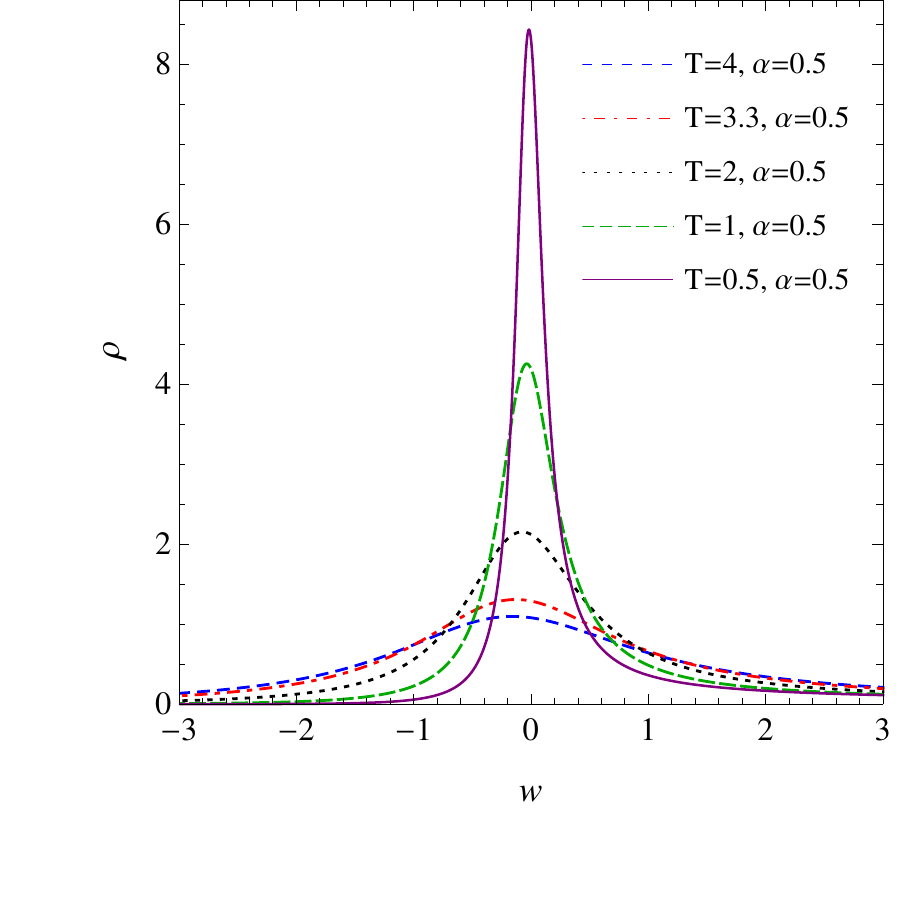}\\[-2em]
  \hspace*{0cm}a.)\hspace*{7cm}b.)
  \caption{The coupling constant dependence of the spectral function
    in the 2PI approximation (a.) at fixed temperature $T=1$, and (b.)
    at fixed fixed coupling value, $\alpha=0.5$. The curves widen with
    growing coupling, and growing temperature.}
  \label{fig:varaandT}
\end{figure}

By applying the algorithm described in Section III/B we can obtain the
spectral function derived from the 2PI approximation for the theory,
using Eq.(\ref{u0}) as the self energy input. On
Fig.~\ref{fig:varaandT} we can see the spectral function for different
coupling values and for different temperatures. The spectrum exhibits
a pole, its width is growing with increasing coupling constant and with
increasing temperature.

In the Dyson-Schwinger approach the exact spectral function can be derived
in a closed form (at least in the the zero velocity case). We wish to
compare the 2PI results to our analytical expression obtained in
\cite{Jakovac:2013aa}:
\begin{eqnarray}\label{exact}
\rho(w)=\frac{N_{\alpha}\beta\sin{(\alpha)} e^{\beta w/2}}{\cosh(\beta w)-\cos(\alpha)}\frac{1}{\left|\Gamma\left(1+\frac{\alpha}{2\pi}+i \frac{\beta w }{2 \pi}\right) \right|^2}.
\end{eqnarray}
Here we used the notation $w=p_0-m$ again and $N_{\alpha}$ is a
normalization factor. Both for the 2PI approximation and the D-S
calculation we assumed a normalization prescription, which assigned by
$\int_{w} \rho=1$ sum rule.

\begin{figure}[h!]
\centering
  \includegraphics[width=6cm]{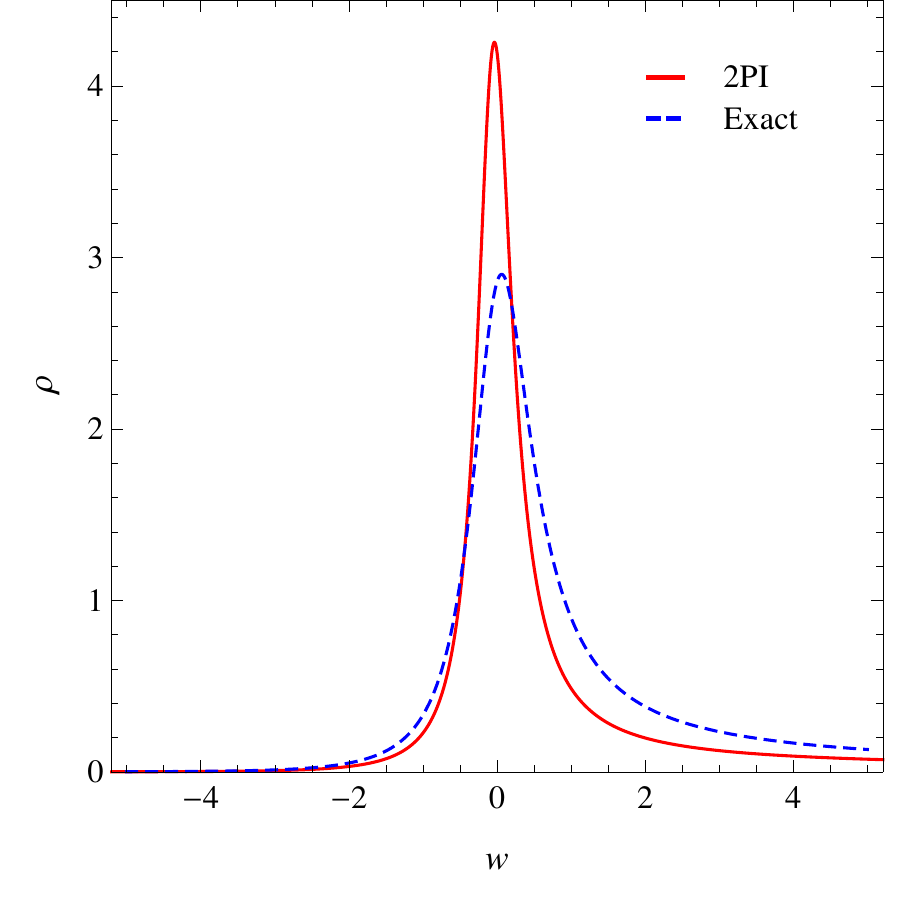}
  \caption{Comparing the 2PI resummed spectral function to the exact
    one. The solid red line is obtained from the 2PI resummation while
    the dashed blue is the exact spectral function. Both of them are
    at T=1 and the couplings are $\alpha_{ex}=\alpha_{2PI}=0.5$.}
  \label{2PIvsEx}
\end{figure}

To check the quality of the 2PI approximation, we can compare the
resulting spectral function with the exact one. The comparison can be seen on
Fig.(\ref{2PIvsEx}).
We can see immediately that the two spectra are not very similar. The
reason is, as we discussed in the Introduction, that the 2PI
approximation do not sum up all the diagrams, in particular the
coupling constant corrections. To improve the 2PI calculation,
therefore, we can try to take into account the resummation of these
diagrams effectively in a renormalization group inspired way, as a
temperature dependent coupling constant. We should use a
nonperturbative matching procedure, and choose that value of
$\alpha_{2PI}$ which reproduces the exact result the most
accurately. For a perfect matching not only the coupling constant, but
also the higher point functions should also be resummed. But we may
hope that the most important effect comes from the relevant couplings,
in this case from $\alpha_{2PI}$.

\begin{figure}[ht]
\begin{subfigure}{.4\textwidth}
  \centering
  \vspace*{0.5em}
  \includegraphics[height=5.85cm]{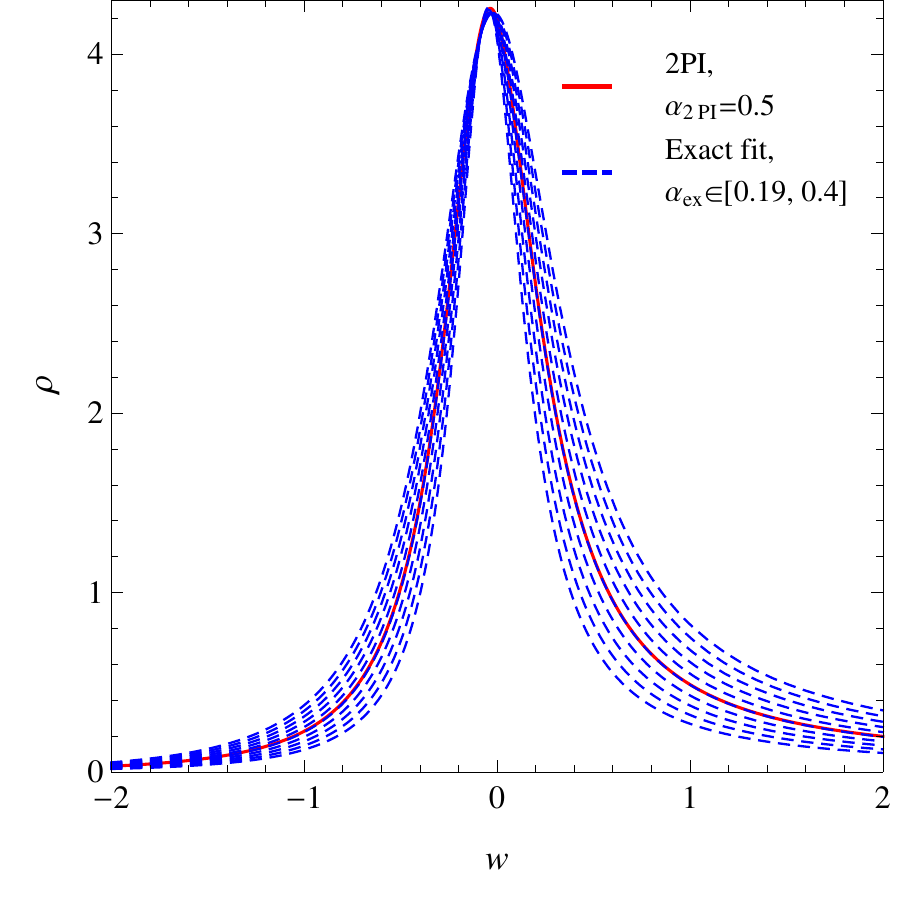}
  \caption{}
\label{plotarmy1}
\end{subfigure}
\begin{subfigure}{.4\textwidth}
\centering
  \includegraphics[height=6cm]{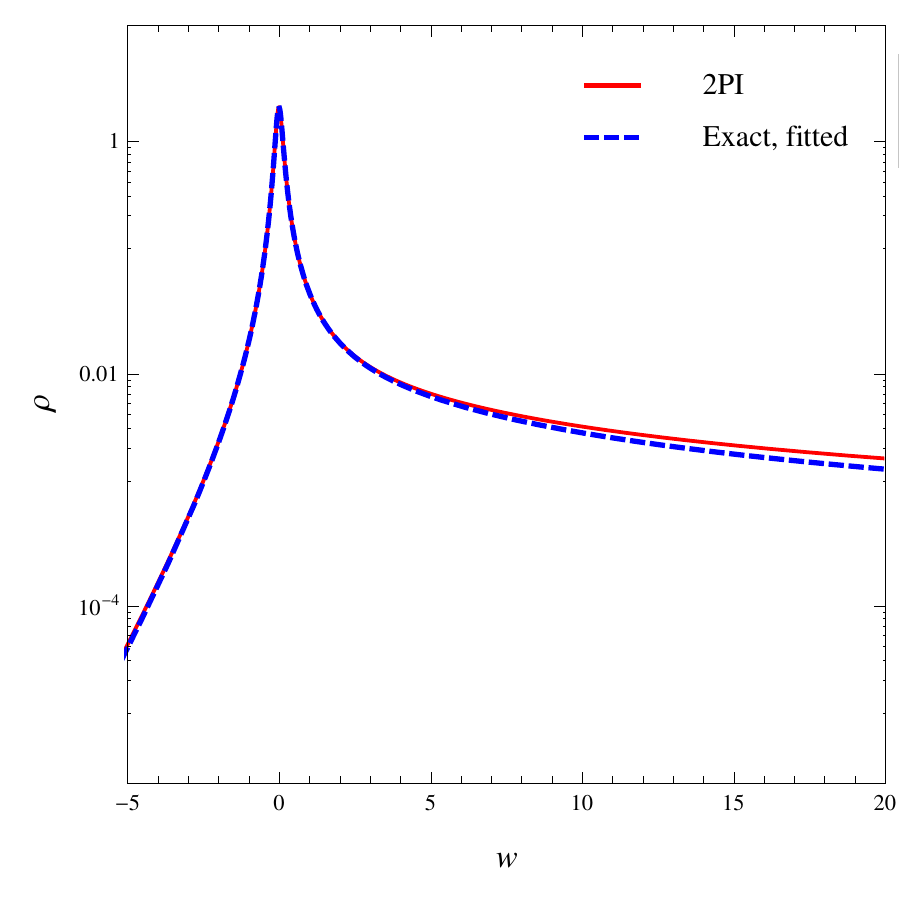}
  \caption{}
\label{ExfitLog}
\end{subfigure}
\caption{The fitting of the exact spectral function on the 2PI
  spectrum in linear (a.) and logarithmic (b.) plot. We can see an
  exact match at the peak and a small deviation in the
  asymptotics. The fit yields $\alpha_{2PI}=0.5$ for
  $\alpha_{ex}=0.293$ at $T=1$.}
\label{fixamp}
\end{figure}

Therefore our strategy will be to find the best, temperature dependent
value of the coupling constant $\alpha_{2PI}$ which yields the best
match between the exact and the 2PI spectral functions. As we can see
on Fig.~\ref{fixamp}, there exist such a value, where the matching is
almost perfect. We can observe that the fit is excellent not just at the
close vicinity of the peak region, but also for much larger momentum
regime, and it can give an account also for the asymmetric form of the
exact spectral function. For asymptotically large momenta we expect
that the two curves do not agree, according to \cite{Jakovac:2011aa},
this can also be observed on Fig.~\ref{fixamp}. This result is a
strong argument in favor of the usability of 2PI technique at finite
temperature also for gauge theories.

Hence we can say that the coupling which takes the value of
$\alpha_{2PI}=0.5$ in the 2PI resummation at $T=1$ is equivalent to an
$\alpha_{ex}=0.293$ in the D-S calculation at the same temperature. One can also conclude that
the vertex corrections (which are absent in the 2PI self-energy
calculations) have a role to modify the value of the renormalized
coupling. In the following we are going to look for a general relation
between $\alpha_{2PI}$ and
$\alpha_{ex}$.

We can repeat the strategy above for different temperatures. In this
way we can determine a relation $\alpha_{2PI}(\alpha_{ex},T)$
(technically it is simpler to obtain $\alpha_{ex}(\alpha_{2PI},T)$ and
invert this relation). This provides the finite temperature
dependence, or finite temperature ``running'' of the 2PI 
coupling constant.

\begin{figure}[htbp]
  \centering
  \hbox{
    \includegraphics[height=8cm]{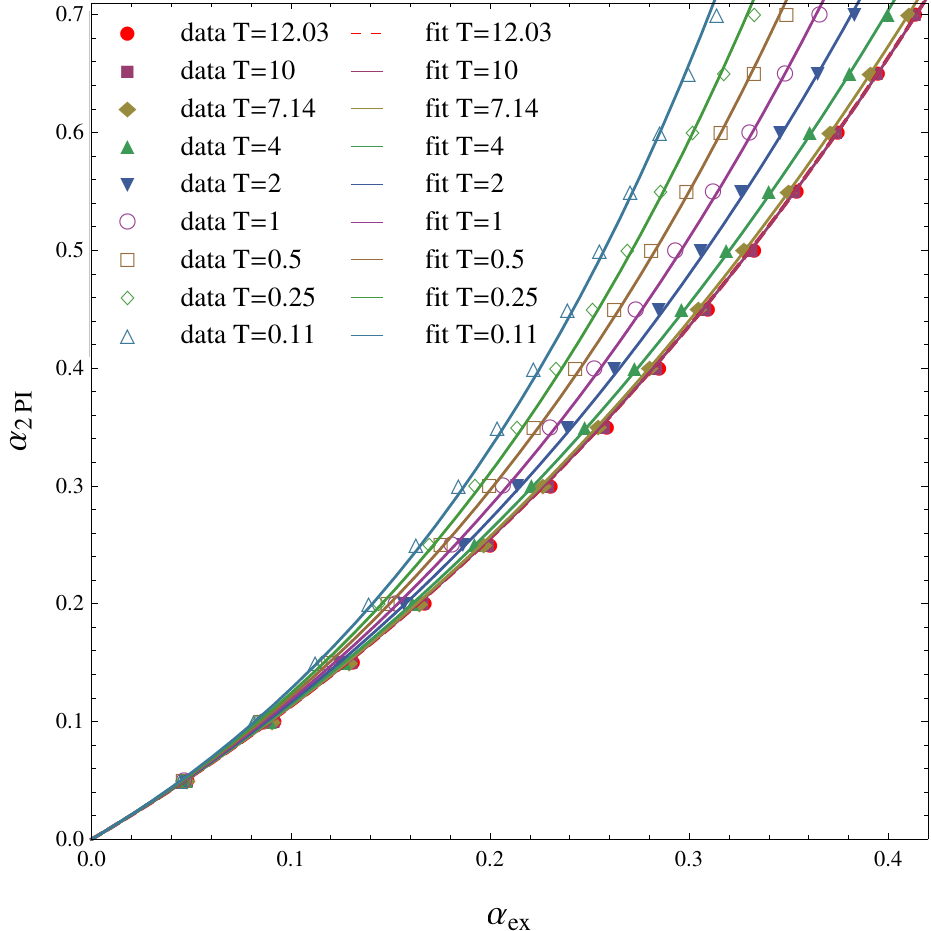}
    \hspace*{1cm}
    \includegraphics[height=8cm]{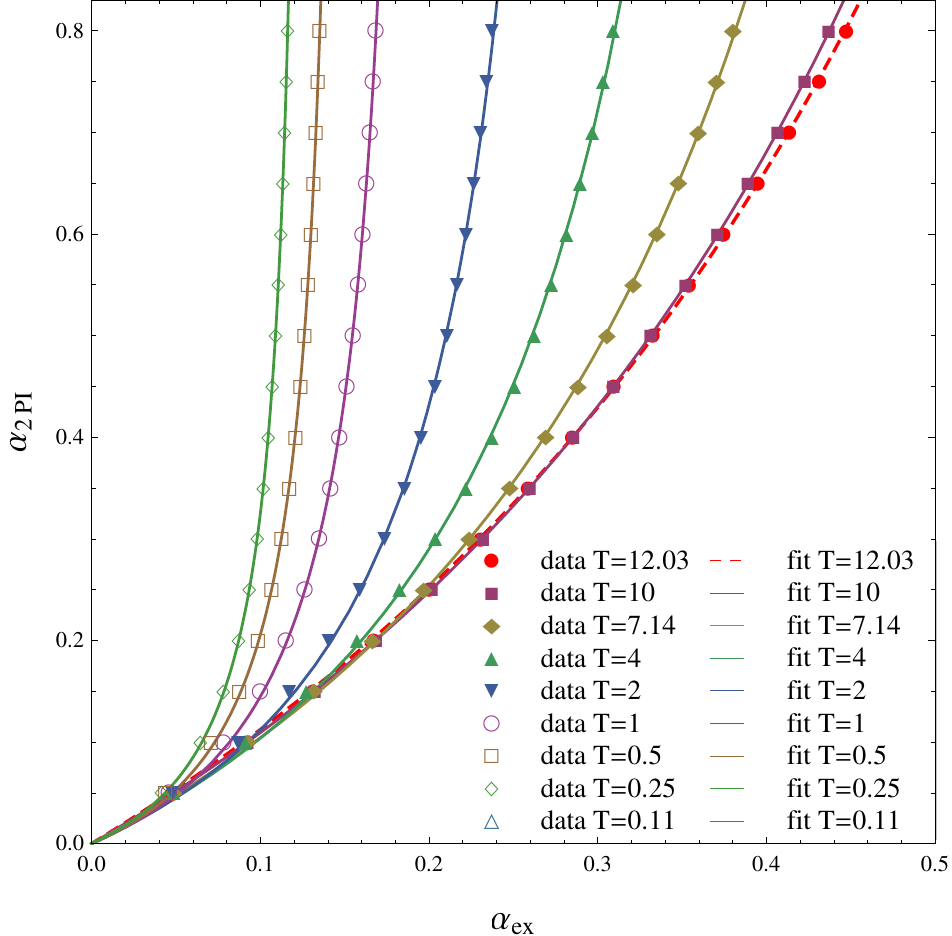}}
  \vspace*{0em}
  \hspace*{0cm}a.)\hspace*{9cm}b.)
  \caption{The relation between the 2PI and the exact coupling at
    $u=0$ for temperatures a.) $T\in[0,12.03]$ and b.)
    $T\in(12.03,\infty)$, respectively. The dashed red line indicates
    the limiting function at $T=12.03$, for details see the text. }
  \label{a_u0}
\end{figure}

We expect that for small couplings the exact and the perturbative
values agree, since the perturbation theory gives $\alpha_{ex} =
\alpha_{2PI} + {\cal O}(\alpha_{2PI}^2)$. This is indeed the case. For
larger couplings, however, the linear relation changes. Interestingly,
we can observe that two different type of functions describe the
relation between the couplings depending on the temperature. The first
type of function which gives the mapping between the to couplings is
valid in the interval $T \in [0,12.03]$. This relation can be obtained
by a one-parameter fit between the 2PI and the exact couplings, namely:
\begin{eqnarray}\label{fit}
\alpha_{2PI}=A_T(e^{\frac{\alpha_{ex}}{A_T}}-1).
\end{eqnarray}   
The result is shown on Fig.~\ref{a_u0}/a., the fit parameters ($A_T$)
are listed in Tables~\ref{tab:fitpar1} and \ref{tab:fitpar2}. From
this relation we immediately see that for small $\alpha_{2PI}$ the
relation of the couplings is linear
\begin{eqnarray}\label{universal}
\alpha_{2PI}\approx \alpha_{ex}+\mathcal{O}\left(\frac{\alpha_{ex}^2}{A_T}\right).
\end{eqnarray}
This tells us that the 2PI and the exact couplings are the same for
the perturbative region, meaning that we can rely on the results
obtained by 2PI calculations in this regime. Thus if we are using
couplings which are in the order of the fine structure constant of QED
($\alpha=1/137$) for instance, one does not even have to worry about
the temperature dependence of Eq.(\ref{fit}).

\begin{table}[htbp]
  \centering
  \begin{tabular}[c]{||l||l|l|l|l|l|l|l|l|l||}
    \hline
    $T$ & $0.11$ & $0.25$ & $0.5$ & $1$ & $2$ & $4$ &
    $7.14$ & $10$ & $12.03$ \cr
    \hline
    $A_T$ & $0.213$ & $0.242$ & $0.27$ & $0.305$ & $0.343$ & $0.384 $
    & $0.414$ & $0.426$ & $0.429$\cr
    \hline
  \end{tabular}
  \caption{The fit parameters in the low temperature case. The error
    of the parameters is $\pm 0.001$}
  \label{tab:fitpar1}
\end{table}

\begin{table}[htbp]
  \centering
  \begin{tabular}[c]{||l||l|l|l|l|l|l|l||}
    \hline
    $T$ & $20$ & $50$ & $100$ & $200$ & $500$ & $1000$ &  $2000$\cr 
    \hline
    $B_T$ & $1.03 \pm 0.003$ & $1.118 \pm 0.008$ & $1.217 \pm 0.014$ &
    $1.312 \pm 0.017$ & $1.381 \pm 0.016$ & $1.38 \pm 0.012$ & $1.3
    \pm 0.006$ \cr
    \hline
    $C_T$ & $1.107\pm 0.001$ & $1.668 \pm 0.025$ & $2.654 \pm 0.054$ &
    $4.241 \pm 0.083$ & $6.937 \pm 0.105$ & $8.951 \pm 0.1$ & $9.991
    \pm 0.055$ \cr
    \hline
  \end{tabular}
  \caption{The fit parameters in the high temperature case.}
  \label{tab:fitpar2}
\end{table}

\begin{figure}[htbp]
  \centering
  \hbox{
    \includegraphics[height=4.9cm]{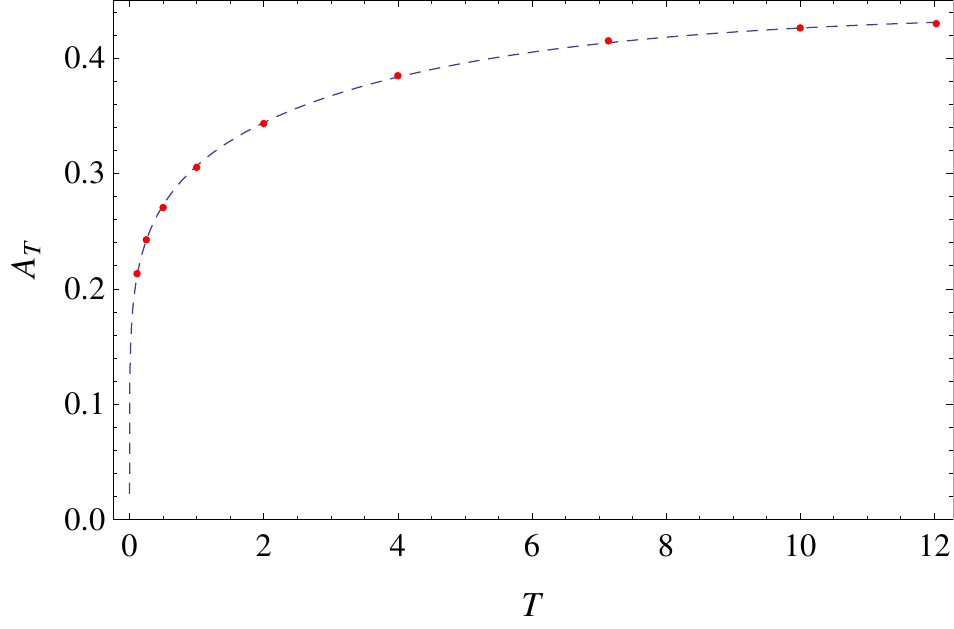}
    \hspace*{1cm}
    \includegraphics[height=5cm]{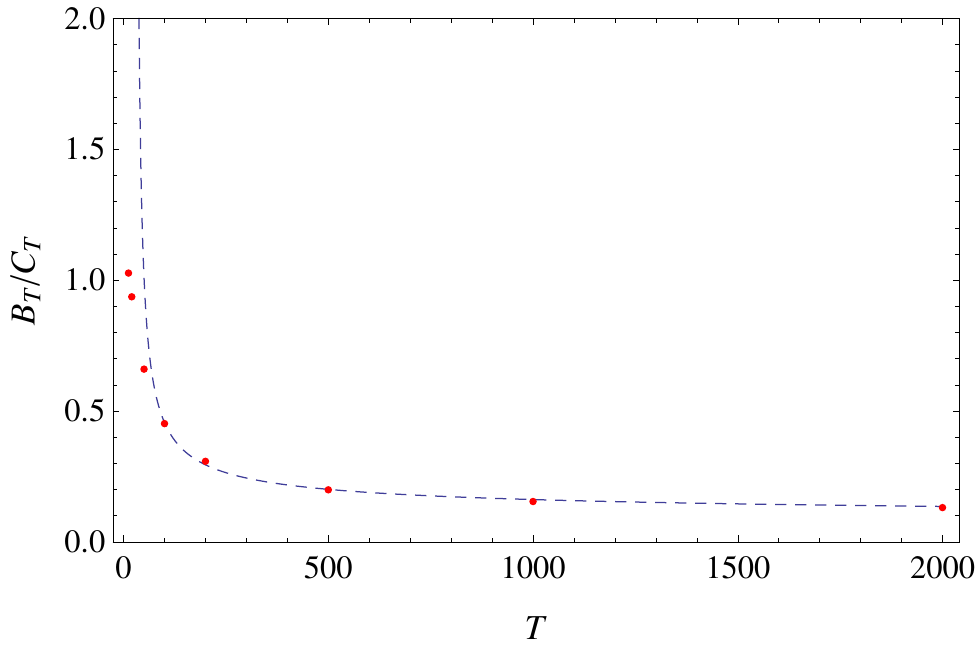}}
  \vspace*{0em}
  \hspace*{-1cm}a.)\hspace*{8.5cm}b.)
  \caption{The running of a.) $A_T$ and b.) $B_T/C_T$ with respect to
    the temperature. This latter quantity is the position of the pole
    (cf. \eqref{fit2}). One can see the best matching on higher
    temperatures.}
  \label{paraA_u0}
\end{figure}

From Eq.(\ref{fit}) it is obvious that the relation depends on the
temperature through the fit parameter $A_T$: this is shown in Fig.\ref{paraA_u0}/a. We can fit
the temperature dependence in the following form:
\begin{eqnarray}\label{tanh}
A_T=a(\tanh{T b})^c,
\end{eqnarray}
where $a=0.438 \pm 0.002$, $b=0.123\pm 0.01$, and $c=0.17 \pm 0.002$. 

Let us consider the zero temperature limit:$\lim\limits_{T\rightarrow
  0} A_T = 0$. This tells us that in the zero temperature limit all
$\alpha_{ex}$ corresponding to any $\alpha_{2PI}$ by Eq.(\ref{fit})
vanishes. To see this it is easier to invert the relation and then
taking the limit, i.e. $\lim_{T\rightarrow
  0}A_{T}\ln(\alpha_{2PI}/A_T+1)=0$. This is consistent with the fact
that at T=0 the coupling drops out from the 2PI propagator
\cite{Jakovac:2011aa}. More precisely at $T=0$ close to the peak:
\begin{equation}
  \label{eq:2piexatT0}
\G_{2PI}(w) \propto \frac{1}{w}\quad\mathrm{, while}\qquad
\G_{ex} (w) \propto \left.\frac{1}{w^{1+\frac{\alpha_{ex}}{\pi}}}\right|_{\alpha_{ex}=0}=\G_{2PI}.
\end{equation}
Therefore the diverging $\alpha_{2PI}/\alpha_{ex}$ relation does not
signal a physical singularity, it just means that in order to match
the exact theory we have to take into account also higher point functions.

The relation in \eqref{tanh} is valid up to the dimensionless
temperature $T=12.03$. Above this temperature the trend of the curves
can be seen in Fig.~\ref{a_u0}/a, namely that they are more and more
shallow for increasing temperature, changes. The
$\alpha_{2PI}(\alpha_{ex})$ curve becomes steeper and steeper as it
can be seen in Fig.~\ref{a_u0}/b. We find for small couplings the
expected universal linear relation
$\alpha_{2PI}=\alpha_{ex}+\dots$. We can also observe that the
$\alpha_{2PI}(\alpha_{ex})$ curves diverge at some limiting value of
$\alpha_{ex}$. This can also be seen from the following fit which
describes the numerically determined curve quite well:
\begin{equation}\label{fit2}
\alpha_{2PI}=\frac{\alpha_{ex}}{B_T-C_T\alpha_{ex}}.
\end{equation} 
The fit parameters can be seen in Table~\ref{tab:fitpar2}. This
function has a pole at $B_{T}/C_{T}$ at each temperature. This is a
temperature dependent quantity, the running of the position of the
pole can be seen on Fig.~\ref{paraA_u0}/b.

Formula \eqref{fit2} can be interpreted from the point of view of the
scale dependence of the coupling constant. For the B-N model the one
loop running is exact \cite{Jakovac:2011aa}, and provides a Landau
pole. The value of the coupling where we find the pole is
$\alpha(\mu_0) = \frac \pi {\ln \mu/\mu_0}$. If we associate $\mu\sim
T$ for high temperatures, this would suggest that the finite
temperature dependence also exhibits a Landau-type pole at
$\alpha_{ex} \sim (\ln f T)^{-1}$.  In fact, a two-parameter fit is
\begin{eqnarray}
\frac{B_T}{C_T}=\frac{d}{\ln(f T),}
\end{eqnarray}
where $d=0.576\pm 0.03$ and $f=0.035\pm 0.003$ describes the finite
temperature behavior for large temperatures. 

\begin{figure}[htbp]
  \centering
  \includegraphics[height=6cm]{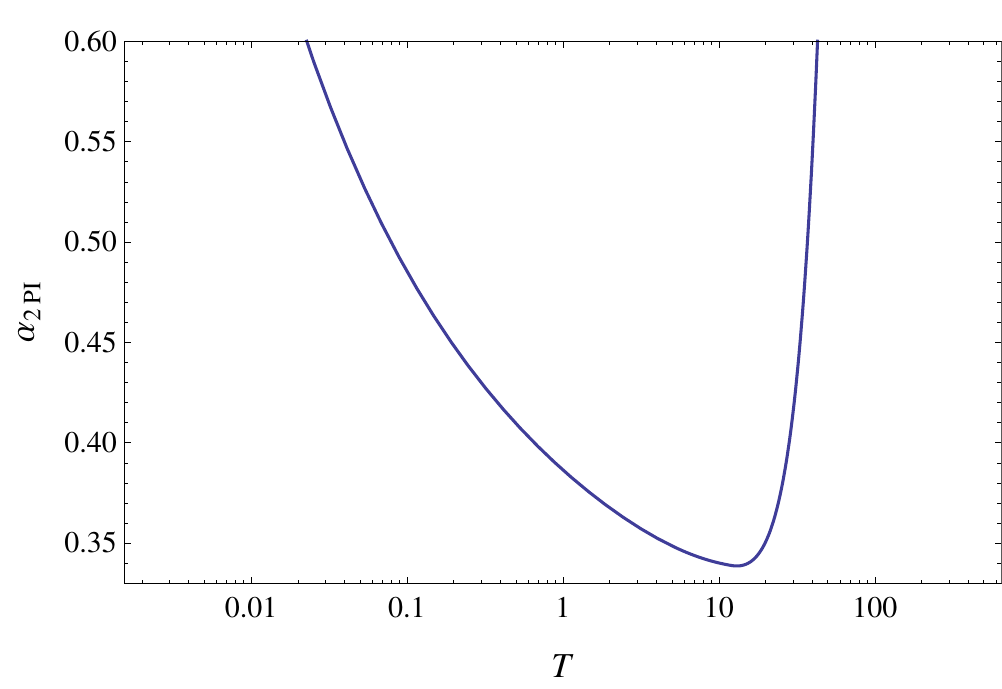}
  \caption{Finite temperature running of $\alpha_{2PI}$ for fixed
    $\alpha_{ex}$. One can observe the high temperature (Landau) pole
    and the $T\to 0$ divergence.}
  \label{fig:alpha2PIrunning}
\end{figure}

The finite temperature running of $\alpha_{2PI}$ for fixed
$\alpha_{ex}$ can be seen in Fig.~\ref{fig:alpha2PIrunning}.
According to our earlier analysis we can identify the following
characteristic features of this running. For small temperatures the
running of the perturbative coupling is determined by the soft IR
physics, the photon cloud. At very small temperatures seemingly we
find a divergence, but this is not a physical singularity, it just
reflects the fact that at zero temperature the 2PI approximation fails
to describe the exact spectrum for any couplings,
cf. \eqref{eq:2piexatT0}. At high temperatures the perturbative
running is the dominant effect with the association $\mu\sim T$. Again
we find a pole there which comes from the Landau pole of the
perturbative running. But again, this singularity is not a physical
one, the exact spectrum is regular for $\alpha_{ex}$ larger than the
pole value. But with 2PI calculation with the original action we
cannot reproduce this result, one would need to take into account
higher point vertices, too. Between the low temperature and high
temperature regimes there is a point where $d\alpha_{2PI}(T)/dT =0$,
in our case this is at the dimensionless temperature value
$T=12.03$. This is a ``fixed point'' of the running and loosely
determines a ``critical temperature'' separating the two physically
different temperature regimes.

\subsection{The finite velocity case}

We can repeat the same analysis for the finite velocity case, too. Since the
findings are very similar to the $u=0$ case, we just shortly overview
the results.

\begin{figure}[htbp]
  \centering
  \includegraphics[height=6cm]{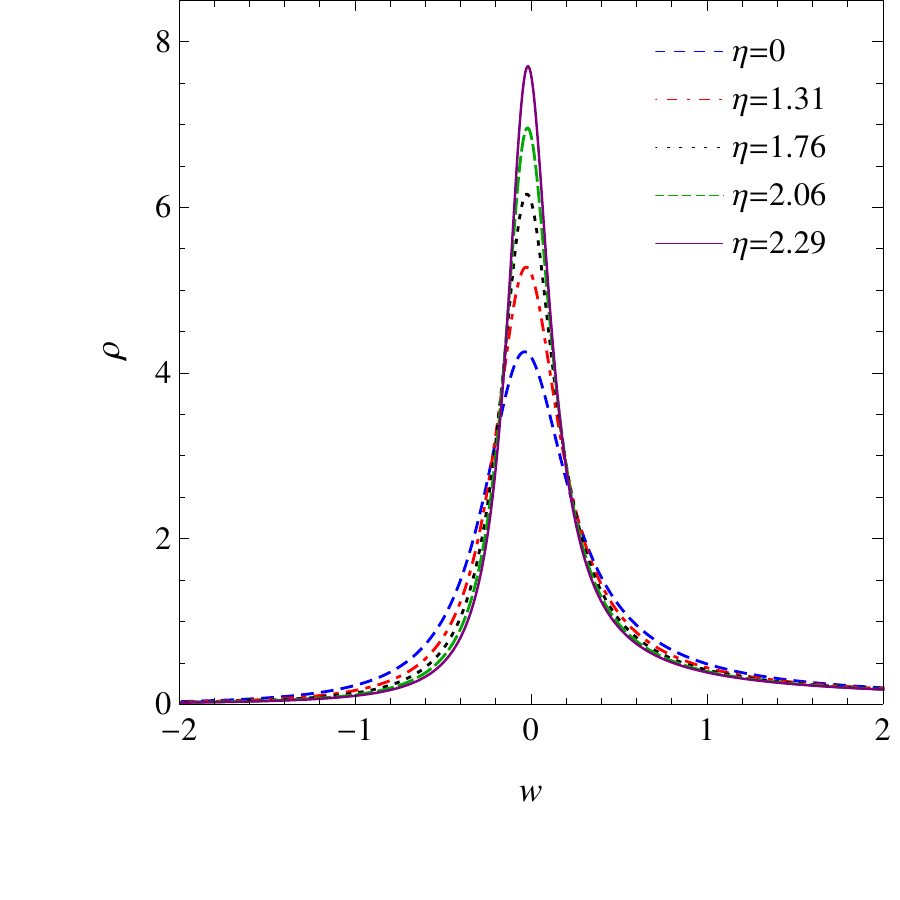}
  \caption{The 2PI spectral functions with different rapidities
    ($\eta=\tanh^{-1}(v)$, where $v=u/u_0$) at fixed temperature $T=1$
    and coupling $\alpha=0.5$. The shrinking of the width can be
    observed as the velocity grows, which is the same effect that we
    had for the exact solution \cite{Jakovac:2013aa}.}
  \label{varumain}
\end{figure}

For the finite velocity case we obtained in our previous article
\cite{Jakovac:2013aa} the following formula in real time:
\begin{eqnarray}\label{finuprod}
\rho(t)\propto z(t)\rho_{u=0}(t;\alpha_{eff}).
\end{eqnarray}
Where we defined an effective coupling which incorporates the
information about the finite velocity
\begin{eqnarray}
\alpha_{eff}=\frac{\alpha u_{0}(1-v^2)}{2v}\ln{\frac{1+v}{1-v}},
\end{eqnarray}
and here $v=\frac{u}{u_0}$. $z(t)$ is a function of time which is defined as:
\begin{eqnarray}
z(t)=\exp\left\{  \frac{u_0(1-v^2)\alpha}{2\pi v} \int\limits_{u_0(1-v)}^{u_0(1+v)} \!\frac{ds}{s^2} \ln\frac{\sinh\pi t Ts}{(\sinh\pi Tt)^s}\right\}.
\end{eqnarray}
In momentum space the product in Eq.(\ref{finuprod}) turns into a
convolution and thus one can derive the finite velocity spectral
function only by using numerics. In the 2PI case we are going to use
the same numerical calculation that we had for the $u=0$ case, the
only difference is that this time we use the formula in
Eq.(\ref{ufin}) for the discontinuity of the self energy. The spectral
functions obtained from 2PI for different $u>0$, but fixed temperature
and coupling constant, can be seen on Fig.(\ref{varumain}).

\begin{figure}[htbp]
  \centering
  \hbox{
    \includegraphics[height=8cm]{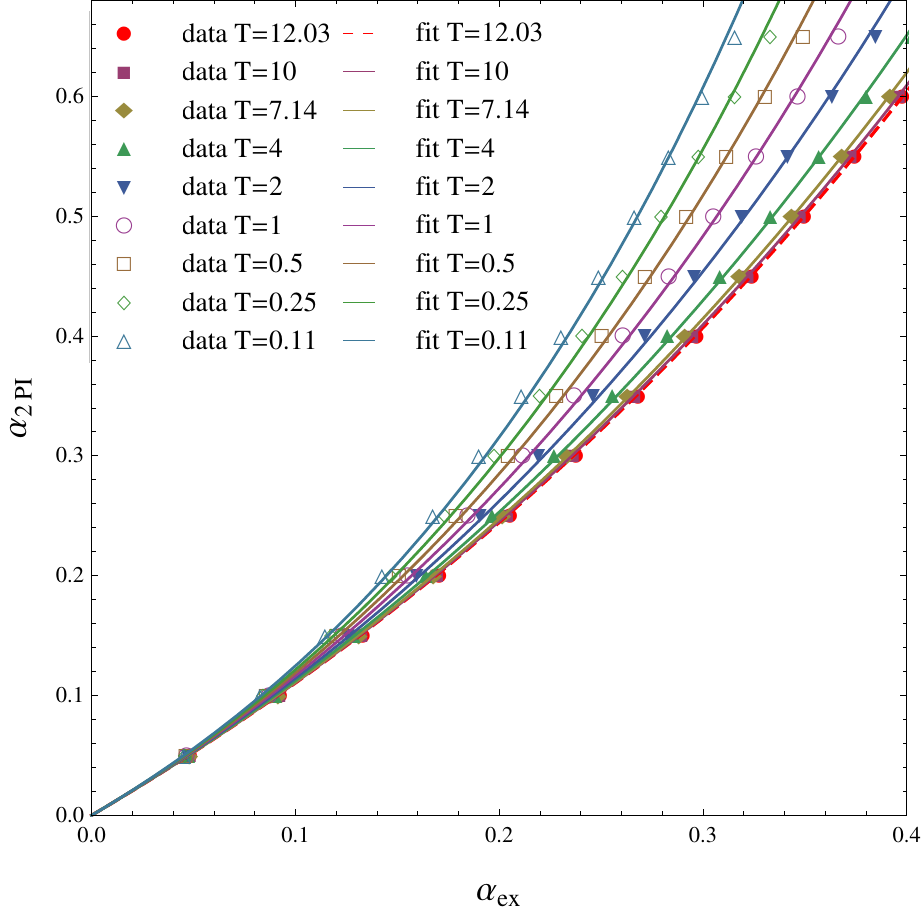}
    \hspace*{1cm}
    \includegraphics[height=8cm]{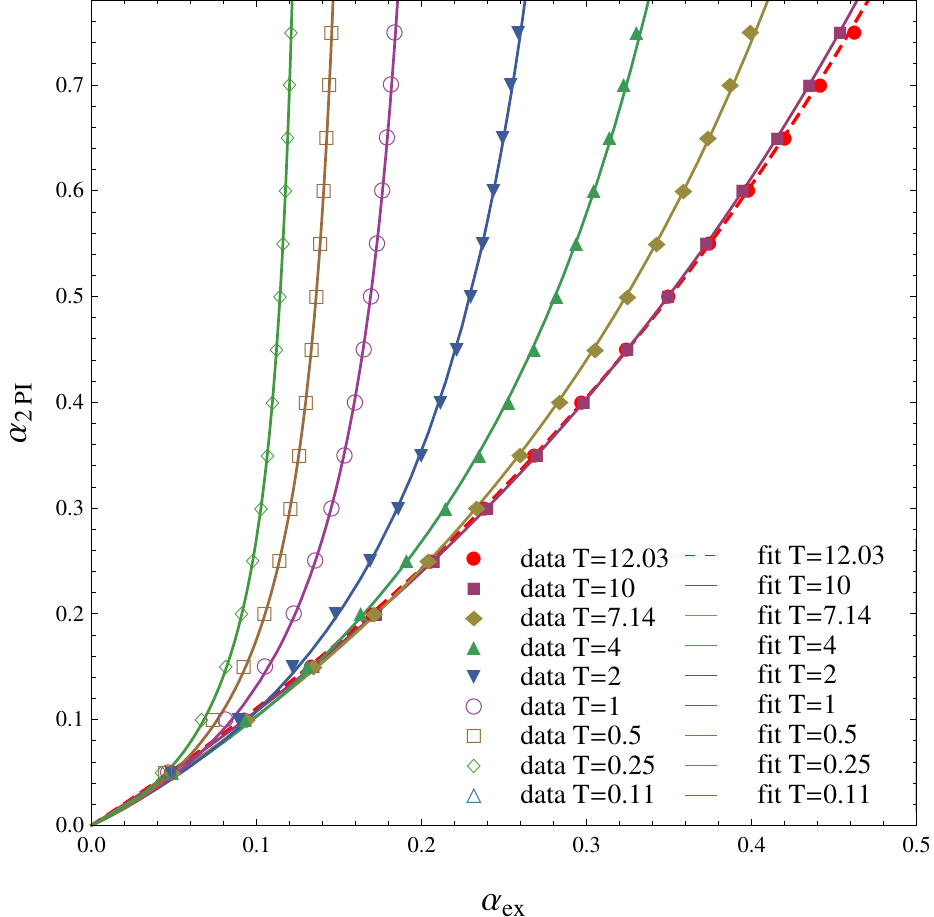}}
  \vspace*{0em}
  \hspace*{0cm}a.)\hspace*{9cm}b.)
  \caption{The relation between the 2PI and the exact coupling for
    $u=\sqrt{3}$ at temperatures a.) $T\in[0,12.03]$ and b.)
    $T\in(12.03,\infty)$, respectively. The dashed red line indicates
    the limiting function at $T=12.03$, for details see the text.}
  \label{a_ufin}
\end{figure}

To fit the $u>0$ spectral functions we are applying exactly the same
procedure that we used for the $u=0$ case. For this purpose we choose
the value $u=\sqrt{3}$ (or $v=\sqrt{3}/2$). In Fig.(\ref{a_ufin}) we
can find the relation between the 2PI and the exact couplings and in
Table~\ref{tab:fitpar1ufin},~\ref{tab:fitpar2ufin} the corresponding
fit parameters, but this time for $u=\sqrt{3}$. For the given finite
$u$ we have almost the same picture that we had for the $u=0$ case,
just the fit parameters, $A_T, B_T$ and $C_T$, are
different. Interestingly the threshold temperature stayed at $T=12.03$
but the running of the parameter as a function of the temperature is
slightly modified. Now we have for $A_T=a\tanh(b A_T)^c$, where this
time $a=0.55\pm 0.01$, $b=0.075\pm 0.01$ and $c=0.183\pm 0.004$. For
the running of the pole we have ($B_T/C_T=d/\ln(fT)$) $d=0.623 \pm
0.04$ and $f=0.032\pm 0.003$.

\begin{table}[htbp]
  \centering
  \begin{tabular}[c]{||l||l|l|l|l|l|l|l|l|l||}
    \hline
    $T$ & $0.11$ & $0.25$ & $0.5$ & $1$ & $2$ & $4$ &
    $7.14$ & $10$ & $12.03$ \cr
    \hline
    $A_T$ & $0.235 \pm 0.002$ & $0.266 \pm 0.002$ & $0.298 \pm 0.002$ & $0.338 \pm 0.002$ & $0.386 \pm 0.002$ & $ 0.442  \pm 0.002 $ & $0.488 \pm 0.002$ & $0.507 \pm 0.001$ & $0.517 \pm 0.001$\cr
    \hline
  \end{tabular}
  \caption{The fit parameters in the low temperature case for $u=\sqrt{3}$.}
  \label{tab:fitpar1ufin}
\end{table}

\begin{table}[htbp]
  \centering
  \begin{tabular}[c]{||l||l|l|l|l|l|l|l||}
    \hline
    $T$ & $20$ & $50$ & $100$ & $200$ & $500$ & $1000$ &
    $2000$\cr 
    \hline
    $B_T$ & $1.016 \pm 0.001$ & $1.108 \pm 0.007$ & $1.2 \pm 0.014$ &
    $1.29 \pm 0.017$ & $1.371 \pm 0.0175$ & $1.389 \pm 0.0145$ & $1.35 \pm 0.009$ \cr
    \hline
    $C_T$ & $0.908 \pm 0.002$ & $1.422 \pm 0.023$ & $=2.275 \pm 0.048$ &
    $3.629 \pm 0.075$ & $6.113 \pm 0.102$ & $8.216 \pm 0.105$ & $9.8318 \pm 0.079$ \cr
    \hline
  \end{tabular}
  \caption{The fit parameters in the high temperature case for $u=\sqrt{3}$.}
  \label{tab:fitpar2ufin}
\end{table}

\section{Conclusion}

We gave a numerical implementation of the 2PI resummation for the
fermionic spectral function in the Bloch-Nordsieck model at non-zero
temperature. In our former paper \cite{Jakovac:2013aa} we showed a
derivation of the exact spectral function in an analytic way and
obtained a closed form. Hence, this analytic formula provides us a
good basis point in the benchmarking of the 2PI. The 2PI technique,
being an approximation, cannot provide us a full solution, but we can
still compare it to the exact result.

Our first main result is that the 2PI approximation works excellently
at finite temperatures, the spectrum coming from the 2PI approximation
could be fitted to the exact spectrum with very good accuracy. The two
curves could be fitted into each other not just in the vicinity of the
peak, but also for much larger momentum interval. This demonstrates
that the 2PI resummation is in fact a physically appropriate
approximation for gauge theories, too.

Nevertheless, the 2PI and the exact results could be fitted to each
other after properly choosing the 2PI coupling
$\alpha_{2PI}(\alpha_{ex},T)$ as a function of the coupling of the
exact formula ($\alpha_{ex}$) and the temperature. For a fixed
$\alpha_{ex}$ this describes a temperature dependent ``running''
coupling constant. Our second main result is to provide this function
for the B-N model.

This temperature dependence has two distinct regimes for small and
large temperatures. At small temperatures the deep IR physics dominate
the running, the corresponding $\alpha_{2PI}(T)$ decreases with the
temperature. For high temperatures the finite temperature running is
compatible with the perturbative scale dependence with the choice
$\mu\sim T$, there $\alpha_{2PI}(T)$ grows with the temperature. At
zero temperature and at some (coupling dependent) high temperatures we
find divergences in $\alpha_{2PI}(T)$, in the high temperature case it
can be associated with the Landau-pole. But none of these poles mean
physical singularity, just the breakdown of the perturbation
theory. Between the two regimes there is a temperature, where the
temperature derivative of $\alpha_{2PI}'(T)=0$. The ``critical
temperature'' of this ``fixed point'' is in dimensionless units
$T=12.03$, this signals the limiting temperature of the soft and
perturbative domains.

The success of the 2PI method extended by a nonperturbative running of
the coupling constant encourages one to try this strategy also in case
of other (gauge) theories. The basis of the temperature running could
be the matching of a nonperturbatively (eg. in MC simulations)
determined physical quantity. Then, using temperature dependent 2PI
couplings one could perform other calculations, and give predictions
for other, numerically hardly accessible physical quantities. 

\begin{acknowledgments}
  The authors acknowledge useful discussions with M. Horv\'ath,
  A. Patk\'os and Zs. Sz\'ep. The project was supported by the
  Hungarian National Fund OTKA-K104292. This research was also
  supported by the European Union and the State of Hungary,
  co-financed by the European Social Fund in the framework of
  T\'AMOP-4.2.4.A/ 2-11/1-2012-0001 'National Excellence Program'.
\end{acknowledgments}

\end{document}